\documentclass[aps,twocolumn,nofootinbib]{revtex4}

\usepackage{psfrag,graphicx}
\usepackage{dcolumn}
\usepackage{amsmath,amssymb}
\usepackage{bm}
\usepackage{amsthm}
\theoremstyle{plain}
\newcommand{\identity}{\openone}

\newcommand{\proj}[1]{\mbox{$|#1\rangle \!\langle #1 |$}}

\newcommand{\half}{\mbox{$\textstyle \frac{1}{2}$}}
\newcommand{\ket}[1]{\left | #1\right \rangle}
\newcommand{\bra}[1]{\left \langle #1\right |}

\begin{document}

\title{Bounding Fault-Tolerant Thresholds for Purification and Quantum Computation}

\author{Alastair Kay}
\affiliation{Centre for Quantum Computation, Department of Applied
Mathematics and Theoretical Physics, University of Cambridge, Cambridge CB3
0WA, UK}

\date{\today}

\begin{abstract}

In this paper, we place bounds on when it is impossible to purify a noisy two-qubit state if all the gates used in the purification protocol are subject to adversarial local, independent, noise. It is found that the gate operations must be subject to less than $5.3\%$ error. An existing proof that purification is equivalent to error correction is used to show that this bound can also be applied to concatenated error correcting codes in the presence of noisy gates, and hence gives a limit to the tolerable error rate for a fault-tolerant quantum computer formed by concatenation. This is shown to apply also to the case where error detection and post-selection, as proposed by Knill, is used to enhance the threshold. We demonstrate the trade-off between gate/environmentally induced faulty rotations and qubit loss errors.
\end{abstract}


\maketitle

\section{Introduction}

The purification of two-qubit states \cite{bipartite_purification:2,bipartite_purification:3} has proven to be a very important protocol in quantum information, particularly with regard to quantum cryptography, where purification allows one to invoke a monogamy argument to ensure that no eavesdropper can gain any information on the resultant correlations. For the purposes of experiments that might try to implement this part of the protocol, it is important to understand its applicability when the gates are imperfect. To date, the only consideration of such noise has been the numerical evaluation of the effect on a particular protocol \cite{bipartite_purification:1,rohde-2007}. However, given the plethora of different protocols (see e.g.~\cite{bipartite_purification:2,bipartite_purification:3,vollbrecht-2004,hostens}), with different purification regimes and yields, it would be useful to bound the performance of any arbitrary protocol. Following our recent studies of bounds on multipartite purification protocols \cite{kay-2006-8,kay-2006}, we continue this line of investigation by describing regions of parameter space for which fault-tolerant purification of two-qubit states is impossible. For a comprehensive review of purification, see \cite{briegel:07}. For fault-tolerant purification, we have in mind the situation where many copies of a state, which through, for example, distribution and storage protocols, have been subjected to a certain level of noise. A purification protocol then attempts to combine these noisy pairs into a reduced number of higher quality pairs. However, this protocol also introduces a level of noise, and the threshold for this protocol to be useful is where the overall sequence at least maintains the same quality of state.

Similarly, in the field of quantum error correction (QEC) \cite{Ste96c,Ste99a,CS96a}, used to stabilise quantum computations, much work has been devoted to calculating fault-tolerant lower bounds i.e.~if a system is subject to noise at a particular error rate, $\varepsilon$, then the lower bound $\varepsilon_{\text{lower}}$ states that if $\varepsilon<\varepsilon_{\text{lower}}$, a computation can be performed to arbitrary accuracy, although it says nothing conclusive if $\varepsilon>\varepsilon_{\text{lower}}$. This is both theoretical \cite{aharanov:99,Got98b,gottesman:2005,Ste98a,Shor_faults,post_sel_thresh:1,post_sel_thresh:2} and numerical \cite{Kni04a}, with the numerical calculations typically suggesting larger lower bounds than the more exacting theoretical proofs.

Upper bounds to the fault-tolerant threshold have also been considered \cite{harrow:03,razborov,falk:06}. These specify a rate $\varepsilon_{\text{upper}}>\varepsilon_{\text{lower}}$ such that if $\varepsilon>\varepsilon_{\text{upper}}$, arbitrarily accurate quantum computation is certainly impossible. Existing upper bounds are surprisingly weak. For example, the strongest general bound is $45\%$ \cite{falk:06}, derived assuming that the majority operations are perfect, and only a single gate is noisy. Other approaches, with more restrictive assumptions, have been able to reduce the bound as far as $15\%$, but only for a particular choice of universal gate set \cite{virmani:04}. Here, we use the equivalence of error correction and one-way purification \cite{bipartite_purification:4} to develop bounds on purification which are applicable to a particular model of fault-tolerance, where a series of error correcting codes are concatenated (cQEC). However, we make no assumption about the gate set, and also assume that noise affects all gates equally. This results in a much tighter bound of approximately $5.3\%$. The calculation of $\varepsilon_{\text{lower}}$ for a cQEC scheme typically proceeds by showing that within the hierarchy of codes, if $\varepsilon<\varepsilon_{\text{lower}}$, from one level of concatenation to the next, the error is reduced. We take the reverse approach of finding the regime where this cannot happen.

The cQEC scheme is used in all but the most recent fault-tolerant lower bound calculations. The idea behind these newer calculations \cite{Kni04a} is that in cQEC scenarios, most gates can be performed transversally, and are comparatively simple to implement. There is then a smaller set of gates, often just one, whose implementation is much more involved and is the major contributor to the fault-tolerant threshold. Often the best way to apply these gates is to first create an ancilla state, and then interact this with the state in question. Instead of directly producing the ancilla, Knill proposes a massively parallel off-line preparation of ancillas, which can be tested with the help of concatenated error detecting codes (cQED) to verify if they have been correctly prepared \cite{Kni04a}. This potentially allows ancillas which are of too low a fidelity to be increased above the lower bound for fault-tolerant cQEC schemes. As with the cQEC scheme, there are massive overheads in terms of the number of qubits required to come close to the calculated lower bounds. Given the two step decomposition, our bounds work automatically on the main cQEC scheme, and by extending the results of \cite{bipartite_purification:4}, we can adapt our bounds so that they also apply in to the off-line preparation component, and, although the results that we produce in this context are weaker than those of \cite{virmani:04}, they are capable of elucidating a variety of trade-offs between different noise rates, such as any asymmetry in gate error rates, and also incorporating the loss of qubits..

In general, for both upper and lower bound calculations, one restricts to a local noise model. For some calculations, such as \cite{gottesman:2005,AKP}, more general noise models can be considered, and, while we primarily concentrate on local noise, we will discuss the potential for extension of our results to these noise models.

\section{Noisy Purification of Two-qubit States}

We shall start by considering bounds on the purification regime of many identical, independently noisy, copies of a two-qubit state shared between Alice and Bob, where they apply local operations (which are also noisy) in an attempt to generate a single, more pure copy.
There are two different
error parameters which are important here -- the initial probability of error,
$p$, on the two-qubit state to be purified, and the probability of an error when
implementing a gate, $q$. In general, the derived bounds will depend on both
of these probabilities.

The four Bell states
\begin{eqnarray}
\ket{\phi^{\pm}}&=&\frac{1}{\sqrt{2}}(\ket{00}\pm\ket{11})	\nonumber\\
\ket{\psi^{\pm}}&=&\frac{1}{\sqrt{2}}(\ket{01}\pm\ket{10})	\nonumber
\end{eqnarray}
form a basis for two-qubit states, and can be inter-converted by the application of the Pauli gates $X$, $Y$ and $Z$. To simplify the analysis, we shall consider the density matrix $\rho$ of the state that we wish to purify to be diagonal in the Bell basis. We shall take it to have fidelity $F$,
$$
F=\max_{U_1,U_2}\bra{\psi^-}(U_1\otimes U_2)\rho(U_1\otimes U_2)^\dagger\ket{\psi^-},
$$
and that, by default, we incorporate the required local unitaries $U_1$ and $U_2$ into the basis, such that $F=\bra{\psi^-}\rho\ket{\psi^-}$. Indeed, any two-qubit state can
be locally converted into a Bell-diagonal state without changing its
fidelity~\cite{bipartite_purification:3}, although this may cause some loss of entanglement. Consequently, we must argue why this restriction to Bell diagonal states is justified, and this will become apparent in our discussion of the particular (adversarial) noise model that we choose. For the moment, let us simply observe that Pauli errors acting on a Bell-diagonal state cause it to remain Bell-diagonal, and that the state towards which we wish to purify is $\rho=\proj{\psi^-}$. In correspondence with the two error probabilities $p$ and $q$, we shall discuss two different fidelities, $F_p$ and $F_q$. $F_p$ is the initial fidelity of the state to be purified. During the purification, we do not
know what the intermediate outcomes of the optimal protocol are. Whatever the operations performed, the final gates
are faulty with errors parametrised by the probability $q$ on
each qubit, and there is no way that this can be compensated for. Hence, during some intermediate step of the process, instead of the fidelity being $F$, it is reduced to $F'$. The fidelity $F_q$ is the fidelity of the state if it would have been perfect except for this final error sequence, i.e.~if $F=1$, $F'=F_q$.

The fidelity of Bell diagonal states is related to the entropy of formation $S(F)$ \cite{wootters:97} by
$$S(F)=H\left(\half+\sqrt{F(1-F)}\right),$$
where $H(x)=-x\log_2(x)-(1-x)\log_2(1-x)$ is the binary entropy. During a purification protocol, we are restricted to acting locally with respect to the tensor product structure of the qubit pairs i.e. for each state $\rho$, the first qubit is held by party Alice, and the second by party Bob. Alice and Bob can perform arbitrary operations on all the qubits they hold, but cannot come together to perform operations on sets of qubits which are jointly held. Under these restrictions, the entropy of formation is a non-increasing quantity, and the aim of the purification protocol is to get as large a value of the entropy as possible between a single pair of qubits, which will involve removing the other qubits, which we describe as `measuring out'. At any arbitrary step $i$ in this protocol, we can consider the combination of $n$ pairs of fidelity $\leq F^{(i)}$, with the aim of producing a single pair, so the $n-1$ pairs which will be measured out cannot transfer more than $(n-1)S(F^{(i)})$ to the purified state on average (due to the sub-additivity of the entropy of formation), such that the best fidelity that can be obtained is $F^{(i+1)}$, satisfying
\begin{equation}
S(F^{(i+1)})=\min(1,nS(F^{(i)})), \label{eqn:combi}
\end{equation}
where the starting point is $F^{(0)}=F_p$.
The errors at the end of the process reduce the entropy to some $S(F')$, where $F'=F_q$ if $S(F^{(i)})=1$. Purification must be impossible if, for all steps of the procedure, the final state is less pure than an initial state
\begin{equation}
F'<F_p.    \label{eqn:conda}
\end{equation}
Recall the context of this single purification step; we are not simply performing the purification with $n$ states, trying to produce a single copy. Instead, this occurs many times in parallel, and the procedure is repeated many times with the outputs, eventually leaving many purified Bell states\footnote{In the same way that in a concatenated fault-tolerant scheme we first error correct blocks of qubits to get more accurate encoded qubits, which we then error correct and so on, eventually leaving sufficient qubits on which to perform a computation}. This is important, because it reminds us that we are interested in the asymptotic rate (for fixed $n$), which notably coincides with the definition of the entropy of formation\footnote{and the way in which we will show the asymptotic equivalence of purification and error correction.}. As a result, probabilistic operations, filterings, encodings in subspaces etc.~can never exceed the above bound, even though on a single shot basis, it would appear that they can.

\begin{figure}[!t]
\begin{center}
\includegraphics[width=0.4\textwidth]{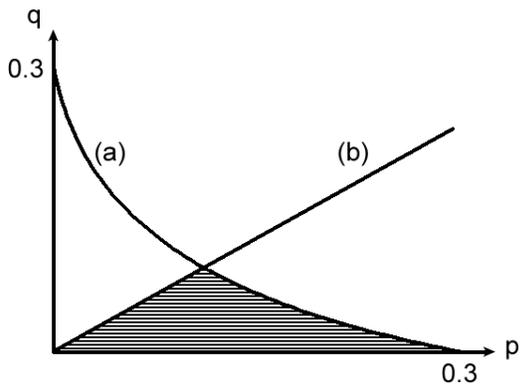}
\end{center}
\caption{The rate of purification for many repetitions of combining $n$ copies of a Bell state with local error probability $p$ to create a single more pure copy goes to zero if we use gates that are faulty with probability $q$ and $(p,q)$ lies outside the
shaded region. The maximum value for $q$ is at the $n$-apex, where the two curves meet. Curve (a) has $F'=F_p$ and (b) has $F_q=F_p$.}
\label{fig:schematic_regime}
\end{figure}

Let us now consider a specific value of $n$, and examine the boundary of the potentially purifiable region. For small $p$ (large $F_p$), $S(F^{(1)})=1$, and hence the boundary is defined by $F_q=F_p$, so as $p$ increases, so does $q$, until $p$ reaches the point that $nS(F_p)=nS(F_q)=1$, which we refer to as the $n$-apex. After this point, the boundary is defined by $F'=F_p$ where $F'<F_q$. Without a specific error model in mind, the behaviour is perhaps not entirely clear, except to note that at the point where $S(F_p)\rightarrow 0$, $q\rightarrow 0$. This region is plotted schematically in Fig.~\ref{fig:schematic_regime}. For our fault-tolerant threshold, we want to find the maximum value of $q$ that we can use, and this must be given by the value on an $n$-apex.

In order to provide some numbers for the error probabilities, it is necessary to assume a particular noise model. We choose to select a noise model which acts independently on each qubit with probability $p$, with noise that is chosen adversarially i.e.~the choice out of all possible options that makes it as difficult as possible to purify. This will be advantageous when examining fault-tolerant error correction as for lower-bound calculations it is typical to assume an adversarial noise model \cite{gottesman:2005}. Consider a Bell-diagonal state $\rho$ on which a rotation $U\otimes\identity$ is applied with probability $p$. We are interested in what rotation is the most destructive with respect to the entanglement in the state. Let
$$
\tilde\rho=(1-p)\rho+p(U\otimes\identity)\rho(U^\dagger\otimes\identity)
$$
and $\tilde\rho_{bd}$ be the state $\tilde\rho$ that has been converted into Bell-diagonal form. Since this can always be done with stochastic local operations, without changing the fidelity \cite{bipartite_purification:3} $$S(\tilde\rho)\geq S(\tilde\rho_{bd}).$$
Note that this conversion is a mathematical tool, not a physical conversion, so there is no concern about the conversion being faulty. If $U=e^{-i\theta\vec{n}\cdot\vec{\sigma}}$ for a 3-component real unit-vector $\vec{n}=(n_x,n_y,n_z)$, then the fidelity is as if the Pauli errors $X$, $Y$ and $Z$ occurred with probabilities $pn_x^2\sin^2\theta$, $pn_y^2\sin^2\theta$ and $pn_z^2\sin^2\theta$ respectively (cross terms such as $X\proj{\psi^-}Z=\ket{\phi^-}\bra{\psi^+}$ are not Bell-diagonal). The lower bound is minimised by $\theta=\pi/2$, and by selecting the single most destructive Pauli operator $X$, $Y$ or $Z$. Furthermore, for this choice of error, $\tilde\rho$ is already Bell-diagonal, and so the minimum possible value of the lower bound to the entropy is achieved, so this is the most destructive to the true entanglement, not just the lower bound. We conclude that the most destructive error is a Pauli error. This justifies our restriction to the purification of Bell-diagonal states.

We choose to assume that the initial state is affected by independent errors on each qubit with probability $p$. On one qubit, these errors are $X$ errors, on the other $Z$. This coincides with our adversarial model, giving the smallest value of $F_p=(1-p)^2$, and has the additional interpretation that it corresponds to the more natural dephasing error model \cite{kay-2006-8} on graph states, since the graph-state basis is related to the Bell basis by a Hadamard rotation on one of the qubits.

\begin{figure}[!t]
\begin{center}
\includegraphics[width=0.4\textwidth]{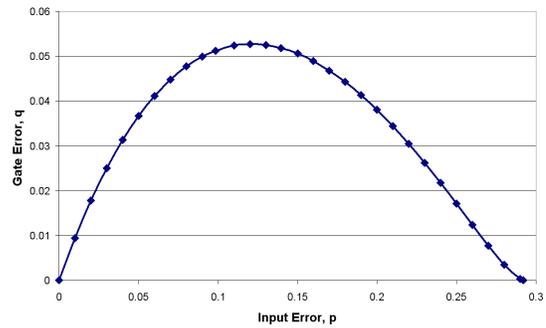}
\end{center}
\caption{A plot of the $n$-apex. The maximum value of $q$ gives the fault-tolerant threshold.}
\label{fig:hump}
\end{figure}

We shall also assume that the noise on the gates, which occurs with probability $q$, is of an adversarial nature i.e.~we can select the local Pauli errors such that they maximally affect the final error probability. Evidently, the worst-case error on qubits that will be measured out is one which propagates through a multi-qubit operation such that it affects the single remaining Bell pair. (For example, consider the controlled-phase gate, where an $X$ rotation on one input propagates to the other qubit as a $Z$-rotation, but $Z$-rotations do not propagate\footnote{A more general way to understand this effect is that we will be making measurements on qubits to remove them. If an error commutes with the measurement basis, then it will have no effect on the remaining state, but there must always be a basis for which this is not true}.) However, since we've had to carefully select our noise such that it propagates onto the single remaining Bell pair, we are no longer free to choose the noise arbitrarily to maximise the error. Thus, to get a valid upper bound, we must instead select the form of Pauli noise which minimises the error. Continuing the assumption of the independence of error, we assume that errors that occurred on different Bell pairs, and are then propagated to the single pair through the purification protocol remain independent. The noise which minimises the error is one where any pair of errors cancels. There are a number of possible choices, for example $Y$ errors on either qubit, but all such possible solutions are equivalent. Nevertheless, the final error that appears on the pair (not having propagated from other qubits) can still be selected to make the noise as bad as possible. Of the $3^2$ possible combinations of Pauli errors, there are only 3 different fidelities which result, the smallest of which is given by
\begin{eqnarray}
F_q&=&(1-q)^2\left(\sum_{m=0}^{n-1}\binom{2n-2}{2m}q^{2m}(1-q)^{2n-2m-2}\right)	\nonumber\\
&+&q^2\left(\sum_{m=0}^{n-2}\binom{2n-2}{2m+1}q^{2m+1}(1-q)^{2n-2m-3}\right)	\nonumber\\
&=&\half\left(1-2q+2q^2+(1-2q)^{2n-1}\right),	\nonumber
\end{eqnarray}
resulting from an $X$ on one qubit, and a $Z$ on the other. In order to simplify the analysis, we consider only the $n$-apex, where $S(F_p)=1/n$ and $F_q=F_p$, allowing us to first eliminate $n$ and then plot $p$ versus $q$, as shown in Fig.~\ref{fig:hump}. One can simply read off this plot that the maximum value of $q$ occurs between the $2$-apex and the $3$-apex, giving a value of $q=0.053$ and $p=0.12$. We quote here the value at non-integer $n$\footnote{for $n=3$, we would have had $(p,q)=(0.140,0.052)$.} in case some hybrid strategy of choosing at random whether to combine 2 or 3 copies of the initial state enables the realisation of this value. 

A final point to mention is that we can consider any interaction (such as a two-qubit gate followed by measurements) between many copies of $\rho$ to be acting like a teleportation, the aim being to teleport as much entanglement from the pairs which will be measured out to the single remaining copy. As such, the error that propagates to the remaining state is also propagated due to the teleportation. One might worry about the trade-off in potential teleportation ability; if we only teleport some of the entanglement, there is a corresponding reduction in the amount of noise that is transmitted. This analysis can be incorporated into the results depicted in Fig.~\ref{fig:hump}, and the maximum bound is given for the perfect teleportation which our analysis (Eqn.~(\ref{eqn:combi})) has assumed.

The case of quantum repeaters \cite{bipartite_purification:1} appears to be even more restrictive. In the previous case, we considered transmission of perfect Bell pairs through a noisy channel, and then purifying. However, in the case of quantum repeaters, we want to use a chain of Bell pairs created over short distances, and then purified, to create a Bell pair shared over a much longer distance. This can be achieved by, for example, sharing one Bell pair between Alice and Bob, and another between Bob and Charlie. Teleporting Bob's end of the first pair through the second pair results in a pair between Alice and Charlie. However, if the Bell pairs are not pure, teleportation is equivalent to transmission through another noisy channel where, this time, the input state is not pure, but only has fidelity $F_q$. This is readily taken into account in our model by changing $F_p$ to depend on both probabilities $p$ and $q$. Note that at the $n$-apex, this still leaves the condition $S(F_q)=1/n$, so the critical value of $q$ would appear to be unchanged. However, we must also note that $F_p<F_q$, so, in fact, all this condition tells us is that provided our first step does not cause any loss of entanglement, future steps won't either.

In this section, we have shown how one can calculate a restricted region in which purification of two-qubit states may be possible even if all the gates that we use experience local noise of a particular (adversarial) model. Specifically, if the gates are faulty with error probability greater than $5.3\%$, purification (by any generic protocol) is impossible. This error probability falls surprisingly close to the numerical performance analyses of recursive purification \cite{bipartite_purification:1,rohde-2007}, which give critical error probabilities in the region of $3-4\%$, once we have converted between the differing notations. Nevertheless, one should note that these analyses consider depolarising noise, which includes the adversarial model with a much reduced probability, and hence we would expect a greater tolerance to this model.

\subsection{Loss Errors}

In different physical scenarios, different types of error become more or less relevant. So far, we have only considered noisy gates i.e.~if the operations are performed imperfectly, then there is an additional rotation on the state, along with similar effects, such as thermal noise, induced by the environment. However, there are other error scenarios that may be relevant. For example, in many experiments, qubit loss is a considerable problem. This may be due to absorption or scattering in an optics experiment, or due to imperfect trapping in experiments with optical lattices, for example. These errors are very different in nature in that we can identify that they have occurred without risking disturbance of the quantum data, and can act accordingly. One
might ask if there are restrictions that can be imposed on the purification
process which would allow non-trivial statements about fault-tolerant purification
thresholds in the presence of qubit loss. One can envision a trade-off in
error-correction between qubit loss and correcting for other types of error
\cite{Rhode:06,Browne:05}.

\begin{figure}[!t]
\begin{center}
\includegraphics[width=0.4\textwidth]{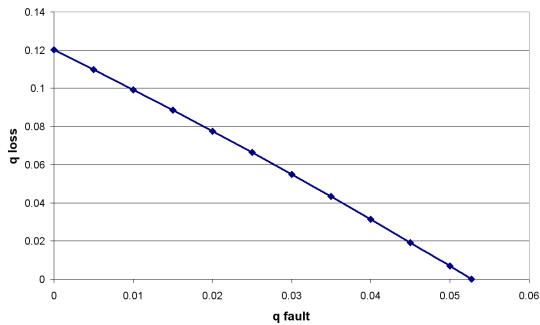}
\end{center}
\caption{The trade-off between gate errors (x) and loss errors (y) in schemes where all gates are equally noisy.}
\label{fig:loss}
\end{figure}

In purification with two-way classical communication (i.e.~where Alice can send measurement results to Bob, and vice versa), loss errors are trivial -- they can be detected. Since they can be detected, Alice and Bob simply exchange information about the location of any lost qubits, and they discard those pairs. Hence purification can proceed as before, and the probabilities of gate faults and of loss are independent. In some situations, it can be interesting to impose further restrictions on the abilities and Alice and Bob. Of particular interest is the case where Alice can communicate with Bob, but not the other way round (this can be thought of as simulating the progression of time), since this will enable us, in the next section, to provide a link between the bounds on fault-tolerant purification and fault-tolerant quantum computation \cite{bipartite_purification:4}. In the previous sections, we didn't enforce this restriction to a uni-directional channel since an upper bound on the performance of an arbitrary purification protocol is also an upper bound on the performance of one restricted to a unidirectional channel. However, when considering loss errors, it behooves us to consider this restriction.

With one-way classical communication, it is also clear how to deal with loss errors -- any loss that Alice detects allows Alice and Bob to discard those pairs, leaving only the pairs where Bob detects a fault. Since he can't inform Alice that there's a fault, his only option is to insert a qubit into the space of the missing one. This may as well be in the maximally mixed state, $\half\identity$. Note that Alice's qubit from the pair is also in this state. We can now update our expressions for $F_p$ and $F_q$,
\begin{eqnarray}
F_p&=&(1-p_\text{loss})(1-p_\text{fault})^2+\frac{p_\text{loss}}{4}	\nonumber\\
F_q&=&\frac{q_\text{l}}{4}+\half(1-2q_\text{f})^{2n-1}(1-q_\text{l})^{2n-1} \nonumber\\
&&+\half(1-q_\text{l})(1-2q_\text{f}+2q_\text{f}^2),	\nonumber
\end{eqnarray}
which arise from the realisation that if there is a loss error, we are given the maximally mixed state, with fidelity $1/4$. If the loss error occurs on one of the qubits to be measured out, an error is transmitted to the remaining qubit with probability $\half$. Otherwise, we are given the state with noise due to the faulty gates. In Fig.~\ref{fig:loss} we have plotted the largest value of $q_\text{loss}$ for a specific value of $q_\text{fault}$ for which fault tolerance could be possible.

\section{Fault-Tolerant Error-Correction}

\subsection{Concatenated Error Correction}

\begin{figure}
\begin{center}
\includegraphics[width=0.3\textwidth]{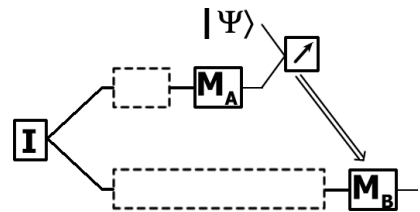}
\end{center}
\caption{A source I prepares Bell states and transmits them through a noisy channel (the dashed boxes) to Alice (top) and Bob (bottom). Alice performs all her operations and sends any measurement results to Bob, who can then, at his leisure, apply his operations to recover the state $\ket{\psi}$ which Alice is teleporting to him.}
\label{fig:1way}
\end{figure}

It has previously been shown that error correction and purification (with one-way classical communication) are equivalent processes \cite{bipartite_purification:4}. This was demonstrated by considering how Alice might transfer an unknown quantum state to Bob when the only quantum data that they share is the output of a noisy channel which distributes noisy Bell pairs to them. The capacity of this channel, $Q_C$, is, by definition, the rate at which Alice can transfer quantum information to Bob using this channel, maximised over all encoding strategies (in particular, error correcting codes) when Alice can send classical information to Bob, but Bob cannot send anything to Alice. Similarly, we can define the one-way purification rate for the Bell pairs distributed by the channel, $D_1$. If the channel distributes $n$ such noisy pairs, Alice and Bob can distil $nD_1$ perfect copies where, again, Alice can send classical data to Bob but not vice versa. This setting is depicted in Fig.~\ref{fig:1way}. In the scenario of quantum computation, the one-way channel has the interpretation that Alice provides the input to a step of the (noisy) computation, and Bob receives the output after the finite time that the step takes. Since Bob cannot communicate backwards in time, the only communication available is a one-way channel.

The proof of equivalence of one-way purification and error correction now follows from considering two different protocols. Firstly, if we take $n$ of the noisy Bell pairs, and distil $nD_1$ pure pairs from them, then Alice can teleport $nD_1$ qubits of information to Bob and therefore at least that much information can be transmitted through the channel, $Q_C\geq D_1$. Secondly, if Alice were to prepare a set of $m$ Bell pairs, and encode one half of each into an error correcting code spanning $n$ qubits, then these can be teleported through $n$ noisy Bell pairs to Bob, who then performs error correction. This ratio $m/n$ is maximised by $Q_C$ and, since this presents a one-way purification protocol, we can say $D_1\geq Q_C$. Combining these two results gives that $D_1=Q_C$. This means that if Alice was trying to share a perfect Bell pair with Bob, they would have the same pair by following either of the two tactics. It should also be clear that if this new Bell pair were also transmitted through a noisy channel, the whole protocol would recurse giving the equivalence between a recursive one-way purification protocol (which is, itself, a one-way purification protocol) and cQEC codes.

We shall now consider the addition of the noise model that we used in the preceding section, and show that, for this model, cQEC codes yield a fault-tolerance which is upper-bounded by the same purification bounds. Recall that this model stated that since we do not know what operations occur during the protocol, which may contain some degree of error detection and/or correction, we only take into account the errors that we know definitely occur and cannot be corrected - the error due to the final gate on each qubit. Hence, this noise model can be considered as transmission through a secondary channel after perfect purification. This same assumption can be made for both purification and concatenated error correction.

\begin{figure}[!t]
\begin{center}
\includegraphics[width=0.3\textwidth]{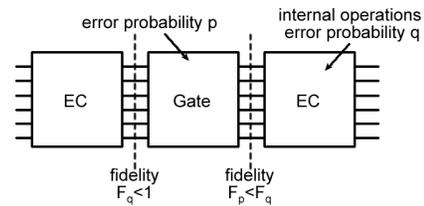}
\end{center}
\caption{In fault-tolerant computation, a gate acting on encoded
qubits is surrounded by error correction (EC) cycles.}
\label{fig:ec_cycles}
\end{figure}

The performance of the protocol can be bounded because the recursive purification procedure is still a global purification procedure, and therefore cannot do better than the values derived in the previous section. In fact, one can argue that it should correspond to the case of quantum repeaters, which we discussed briefly. The reason for this is that in a quantum computation, we have many repeating rounds of correction followed by gates acting on the logical qubits. The output from a particular round of error correction is already faulty with fidelity $F_q$ due to the imperfect operations. It then goes through the logical gate operation, which introduces errors with probability $p$ before starting another round of error correction (see Fig.~\ref{fig:ec_cycles}). The input fidelity at the start of this error correction cycle, $F_p$, must correspond to that of the quantum repeater.

Since an upper bound on two-way purification is also a (possibly weak) upper bound on one-way purification, we conclude that fault-tolerant computation is impossible for a cQEC scheme if local errors occur with probability $>5.3\%$. Note that this made no assumption about the type of gates employed in the scheme, or the QEC code used, simply that all gates are equally noisy. The trade-offs between noise and loss, as depicted in Fig.~\ref{fig:loss} are also applicable.

\subsection{Concatenated Error Detection}

Recently, Knill has proposed a new method of post-selected quantum computation which appears to give tighter threshold bounds \cite{Kni04a,post_sel_thresh:1,post_sel_thresh:2}. In the most popular error correcting codes, such as the Steane 7-qubit code, most of the gates in a universal set are implemented comparatively easily (with transversal operations), and there is typically only a single gate (Toffoli or $\pi/8$) which is responsible for the very small error tolerance of the scheme (because it has to be constructed out of a large number of primitive gate operations), hence the justification of previous works \cite{harrow:03,razborov,falk:06,virmani:04} in considering perfect Clifford operations, and only a single noisy gate. If these gates, which are performed by preparing an ancilla state and then a measurement operation, can be improved, the threshold can be significantly enhanced. Knill's method proceeds in two steps. Firstly, there is an off-line stage, where a series of cQED codes acting on states with error probability below the cQEC threshold are used to prepare the required ancilla states with an error probability greater than the threshold. From there, the second step, the previous cQEC scheme, can proceed.

The discussion of the previous subsection, equating error correcting codes and one-way purification clearly does not apply to the first step, where the QED codes are used. However, we can use similar ideas to show that if purification with two-way classical communication is impossible, so is the error detection scheme. Note that this will mean that the discussion on loss errors will no longer apply to this component. In this step, we consider the specific process of creating the ancilla state, which occurs outside of the standard flow of time for the quantum circuit by virtue of the fact that we can prepare as many copies of the ancilla as we wish, and keep preparing them, until we have found a state that is good enough that we're happy to use it. This gives the intuition for why we should be able to consider two-way communication channels. As such, we now consider our previous setting of purification, except that we will associate the $p$ with the probability of error when producing the ancilla, and that we will then perform some processing on these ancillas using all our other quantum gates, which have probability of error $q$, hoping to produce a better quality ancilla.

\begin{figure}
\begin{center}
\includegraphics[width=0.2\textwidth]{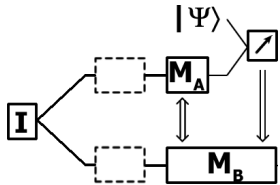}
\end{center}
\caption{A source I prepares Bell states and transmits them through a noisy channel (the dashed boxes) to Alice (top) and Bob (bottom). Alice and Bob can communicate with each other through a two-way classical channel to prepare a perfect Bell pair so that Alice can teleport the state $\ket{\psi}$ to Bob.}
\label{fig:2way}
\end{figure}

We prove the equivalence of two-way purification and error detecting codes by considering the same channel as before, distributing noisy Bell pairs to Alice and Bob. However, they now have two-way classical communication. We define a new channel capacity $Q_D$ as the rate, maximised over all encodings into QED codes, at which Bob receives quantum information from Alice. Clearly this is a very different quantity from before, as Alice and Bob will now be able to detect some errors, and request that information be sent again. Nevertheless, we can follow almost identical arguments to previously. Firstly, we can take $n$ noisy Bell pairs and distil $nD_2$ perfect pairs with a two-way purification protocol, thus allowing transmission of quantum information at a rate $D_2\leq Q_D$. Secondly, Alice can encode $m$ halves of perfect Bell pairs in an $n$-qubit QED code, and teleport them to Bob through the noisy channel. If Bob detects any errors, he can request a new copy be sent. Alice and Bob can therefore share perfect Bell pairs at a rate $Q_D\leq D_2$. So, we have proven that $Q_D=D_2$, and this can be extended as previously to the case of concatenated codes. Thus, the coincidence of numerical results for the recursive purification regime in two-way purification under local depolarising noise \cite{bipartite_purification:1} and the cQED scheme with the same noise model, both functioning in the $3\%$ region, is now revealed as, in essence, the same result. Consequently, we can now examine Fig.~\ref{fig:hump} and examine the trade-offs between the noise that we might be able to tolerate in the Clifford gates ($q$), compared to those of the ancilla preparation ($p$).

For example, if ancilla preparation is more than $12\%$ noisy\footnote{note that this is a measure of the final fidelity, and if a sequence of gates are required to prepare the state, these individual gates need to be higher quality}, and the gates used in the cQED processing have more than $5.3\%$ noise, the fidelity of the ancilla certainly can't be improved. We would need to be able to improve it to the same $5.3\%$ level before it could possibly be input into the main cQEC sequence that is protecting the computation. In this context, these results give bounds that are weaker than those of \cite{virmani:04}, which simply stated that if Clifford gates are perfect, and ancilla preparation is more than $15\%$ noisy, fault-tolerant computation for a particular set of gates is impossible (in our analysis, if we assume that the Clifford gates are perfect, $q=0$, the ancilla cannot be more than $30\%$ noise).

The behaviour of the cQED scheme under loss is exactly the same as the cQEC scheme because within the two stage process, the first stage can tolerate loss -- the losses can be detected and replaced, and it is only the second stage, which is the cQEC stage, which experiences faults due to qubit loss.

\section{Conclusions}

In this paper, we have shown that purification of many copies of a Bell diagonal state is impossible if the gates experience adversarial local noise with a probability greater than $5.3\%$. This bound also applies to fault-tolerant computation using concatenated error correcting codes i.e.~a cQEC or cQED scheme subject to local adversarial noise cannot perform arbitrarily accurate quantum computation if the gates are more noisy than $5.3\%$. While we have assumed that the noise processes that contribute to fidelities $F_p$ and $F_q$ act locally and independently on the qubits, in principle this assumption can be relaxed. All we actually require is that the fidelities of all the states to be purified are uncorrelated. So, for example, this allows $F_p$ to have contributions where the noise is correlated across the two qubits of the pair. In combination with the recent work of Brandao and Eisert \cite{Jens:07}, further analysis of this case may incorporate a large degree of possible error models. Broadly speaking, \cite{Jens:07} shows that if the states are correlated, then the condition of their purification simply reduces to the consideration of the reduced state, and whether it can be purified. Note, however, that an adversarial correlated noise model necessarily incorporates local noise. Thus, our upper bound serves as an upper bound for more general noise models. Incorporating other possibilities can only tighten the bound, in comparison to lower bound calculations where an expanded noise model weakens the bound.

This work is the first time that an upper bound has been calculated where it is assumed that all gates are equally noisy. Moreover, the assumption of noise processes makes it directly comparable to calculations of lower-bounds such as \cite{gottesman:2005}. In fact, our bound is extremely tight with the best-known lower bound, which uses the technique of post-selected computation, a combination of cQED and cQEC schemes. Numerically, this appears to give a threshold of about $3\%$~\cite{Kni04a}, although recent rigorous results suggest a smaller value of $0.1\%$ \cite{post_sel_thresh:1,post_sel_thresh:2}. In contrast, previous works have assumed that just one gate out of a universal set is noisy. In addressing the cQED model, we adapted our derivation to this case, but find that our bound is worse than some existing values.

There are still a number of interesting open questions. Foremost, some of our steps that provide bounds seem quite weak, and it would be worth investigating whether they can be tightened. For example, in the case of error correction, one is restricted to a finite number of copies. What influence does this have on how useful the process may be? Further, one might wonder if we could provide tighter bounds on the performance of a purification protocol with only unidirectional classical communication from Alice to Bob -- when using faulty gates (with the exception of lossy gates, where we have partially taken the condition into account), our bounds are equally applicable to two-way communication, and hence to both the error correcting and error detecting schemes, between which we would expect to see a separation. 

\acknowledgments

The author would like to thank Jiannis Pachos for a critical reading of the manuscript, and Frank Verstraete and Norbert Schuch for useful conversations. Support is provided by Clare College, Cambridge.

\end{document}